\documentclass[aip,rsi,preprint,graphicx]{revtex4-1} 

\usepackage{graphicx}
\usepackage{natbib}
\usepackage[usenames,dvipsnames]{color} 

\draft 

\begin{document}


\title{Insertable system for fast turnaround time microwave experiments in a dilution refrigerator} 



\author{Florian R. Ong}
\email[]{fong@uwaterloo.ca}
\affiliation{Institute for Quantum Computing, Waterloo Institute for Nanotechnology,
and Department of Physics and Astronomy, University of Waterloo, N2L 3G1, Waterloo, Canada}

\author{Jean-Luc Orgiazzi}
\affiliation{Institute for Quantum Computing, Waterloo Institute for Nanotechnology,
and Department of Electrical and Computing Engineering, University of Waterloo, N2L 3G1, Waterloo, Canada}

\author{Arlette de Waard}
\affiliation{Leiden Cryogenics b.v., Kenauweg 11, 2331 BA Leiden, The Netherlands}

\author{Giorgio Frossati}
\affiliation{Leiden Cryogenics b.v., Kenauweg 11, 2331 BA Leiden, The Netherlands}

\author{Adrian Lupascu}
\email[]{alupascu@uwaterloo.ca}
\affiliation{Institute for Quantum Computing, Waterloo Institute for Nanotechnology,
and Department of Physics and Astronomy, University of Waterloo, N2L 3G1, Waterloo, Canada}


\date{\today}

\begin{abstract}
Microwave experiments in dilution refrigerators are a central tool in the field of superconducting quantum circuits and other research areas. This type of experiments relied so far on attaching a device to the mixing chamber of a dilution refrigerator. The minimum turnaround time in this case is a few days as required by cooling down and warming up the entire refrigerator. We developed a new approach, in which a suitable sample holder is attached to a cold-insertable probe and brought in contact with transmission lines permanently mounted inside the cryostat. The total turnaround time is 8 hours if the target temperature is 80 mK. The lowest attainable temperature is 30 mK.  Our system can accommodate up to six transmission lines, with a measurement bandwidth tested between DC and 12 GHz. This bandwidth is limited by low pass components in the setup; we expect the intrinsic bandwidth to be at least 18 GHz. We present our setup, discuss the experimental procedure, and give examples of experiments enabled by this system. This new measurement method will have a major impact on systematic ultra-low temperature studies using microwave signals, including those requiring quantum coherence.

\end{abstract}

\pacs{
07.20.Mc,  
84.40.-x,  
85.35.-p,  
03.67.-a   
05.60.Gg   
}

\maketitle 



\section{Introduction}
A growing variety of experiments requires the combination of ultra-low temperatures (below 100 mK) and the application and detection of electrical signals with bandwidth as large as tens of GHz. These experiments cover a wide area of research, including quantum computing with solid-state devices \cite{clarke-wilhelm,nori-review2011-2}, quantum optics on chip \cite{nori-review2011}, development of quantum limited amplifiers \cite{clerk2010,muck_2001_1,lehnert2008,devoret-JPC2010}, nanoelectromechanical resonators \cite{vandersant2009,oconnell2010,teufel_2011_Cooling}, fundamental transport phenomena in mesoscopic devices \cite{nazarov-QT,schoelkopf_1997_freqdepshotnoise,gabelli_2004_HBTMeso}, and broadband microwave spectroscopy in Corbino geometry \cite{steinbergRSI2012}. The combination of millikelvin temperatures and microwave frequencies arises naturally when studying mesoscopic physics in solid-state systems. On the one hand quantum effects typically become relevant at low temperatures. Examples include collective behaviour in superconductors~\cite{tinkham}, or the increase of the electronic coherence length beyond system size in mesoscopic systems~\cite{imry}. On the other hand microwaves are needed to probe relevant energy scales, such as the plasma frequency in Josephson junctions \cite{clarke-wilhelm}, or the charging energy and Zeeman splitting in semiconducting nanostructures \cite{nazarov-QT,nadjperge2010}. Furthermore,  lowering the temperature $T$ down to a regime where $k_{\rm B} T \ll h \nu$, where $k_B$ is the Bolzmann constant, $h$ is the Planck constant and $h \nu$ is the energy gap from the ground to the first excited state of the quantum system, enables preparation of the ground state and results in optimal coherence~\cite{walls_1995_1}.

The research areas enumerated above require that a mesoscopic device is placed in a dilution refrigerator. Preparation of the cryostat for cooldown (installation of vacuum cans and radiation shields, pumping of large volumes) and the actual cooldown to millikelvin temperature take, depending on the configuration of the system, a time varying between half a day and three days. To warm up the system and then replace the device, an additional time of a few hours to a day is needed. Combining a cold insertable probe developed by Leiden Cryogenics \cite{leiden-website} and a new type of sample connection system we were able to reduce these overhead times dramatically. The turnaround time, defined here as the minimum time needed to measure two successive devices at 80 mK, is reduced to 8 hours. This method has a significant impact on experiments which require multiple device testing.


\section{Experimental setup}
    \label{section-setup}

In this section we present the experimental setup, which builds on a cryogen-free dilution refrigerator type CF-650 and a cold-insertable probe, both available from Leiden Cryogenics \cite{leiden-website}.

\subsection{Dilution refrigerator with cold-insertable probe}
\begin{figure}[htbp]
\includegraphics[scale=0.5]{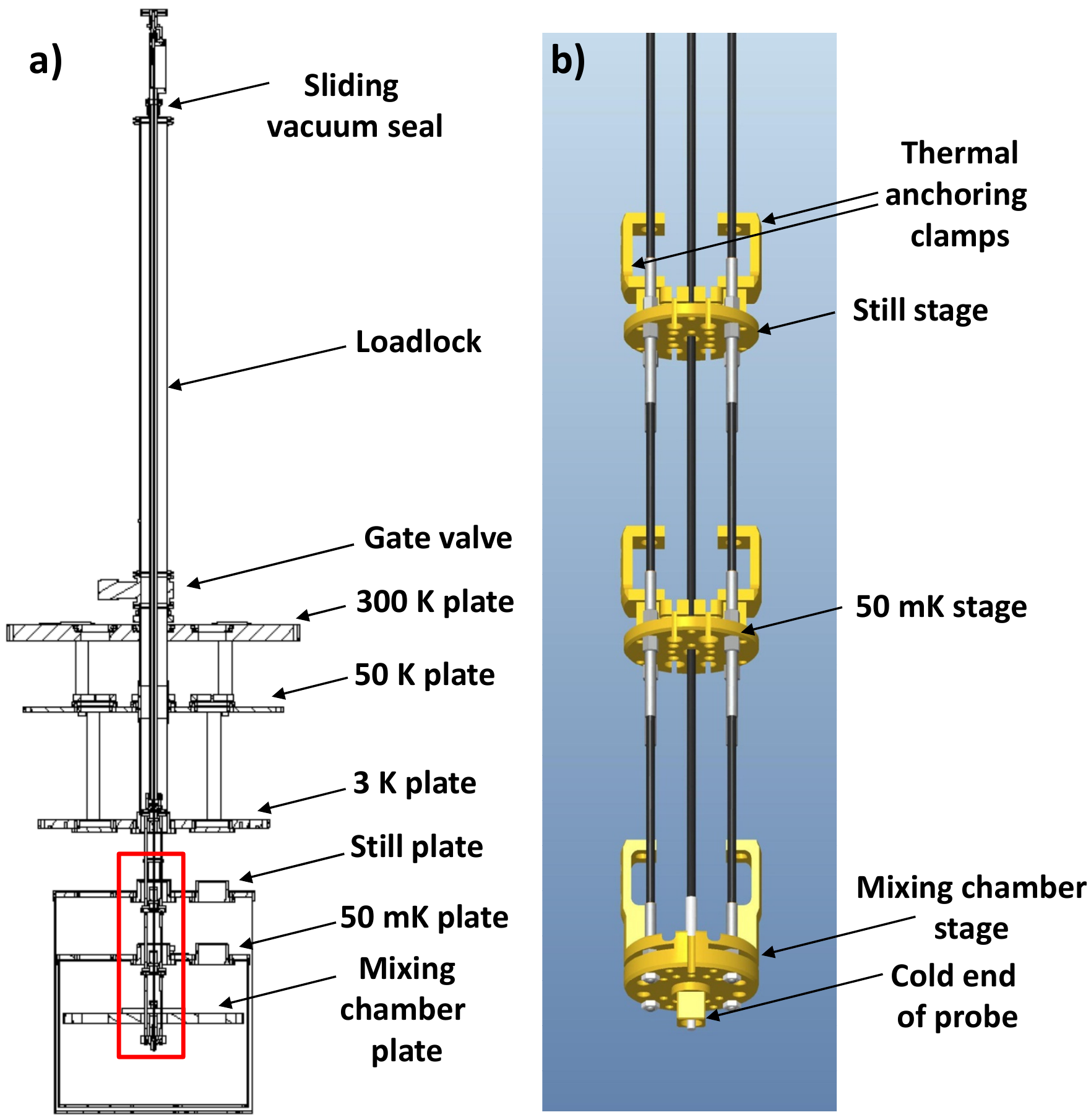}
 \caption{
 \label{fig1}
Section view of the CF-650 refrigerator and its insertable probe. a) General view of the system. b) Close-up view of the low-temperature parts of the probe, corresponding to the red rectangle in panel a).}%
 \end{figure}

The dilution refrigerator CF-650 has the following basic characteristics. The cooling power is 650~$\mu$W for an operation temperature of 120 mK. The base temperature, without experimental wiring installed, is 12 mK.  After the installation of wiring for our experiments, the lowest temperature reached at the mixing chamber is 20 mK.  Three line-of-sight access  ports with a 50 mm diameter run through the inner vacuum can (IVC) from the top plate down to below the mixing chamber. One of them can be used to fit a 2 meter long cold insertable probe (see Fig. \ref{fig1}). The thermalization of the probe is achieved by a mechanism in which anchoring clamps attached to the successive probe stages are brought into contact with the fixed plates of the refrigerator at different temperatures. To establish contact, the anchoring clamps are moved sideways using a knob at the top of the probe.

The Leiden Cryogenics insertable probe comes with a loadlock chamber that can be clamped on any of the three IVC access ports (Fig. \ref{fig1}.a). Apart from thermometry related wires, the probe is fitted in its standard configuration with a set of twelve twisted pair wires which can be used for low-frequency electrical measurements. It is possible to add more wires. However, the addition of transmission lines is only possible to a limited extent. The reason is that a coaxial cable has a diameter of the order of millimiters for reasonably low attenuation at high frequencies. Coaxial connectors, filters and attenuators occupy an even larger space. An even more severe problem occurs for experiments involving low-noise microwave measurements, which require the installation of circulators/isolators and amplifiers. These packaged components have a bulky profile which could not fit in the space allowed by the probe. While a larger probe diameter is possible, such a design would make manipulation more difficult and also would lead to increased heat leakage. An additional problem is the fact that microwave amplifiers dissipate a significant amount of heat (typically 1-100 mW), and therefore are less efficiently thermalized when mounted on the insertable probe than when thermally anchored to one of the cold plates in the refrigerator.

Motivated by the difficulties enumerated above with adding transmission lines on the cold insertable probe we introduce a method for sample insertion explained in the next subsection.

\subsection{Fixture for guided insertion of sample holder and coupling to microwave lines }

\begin{figure}[htbp]
\includegraphics[scale=1]{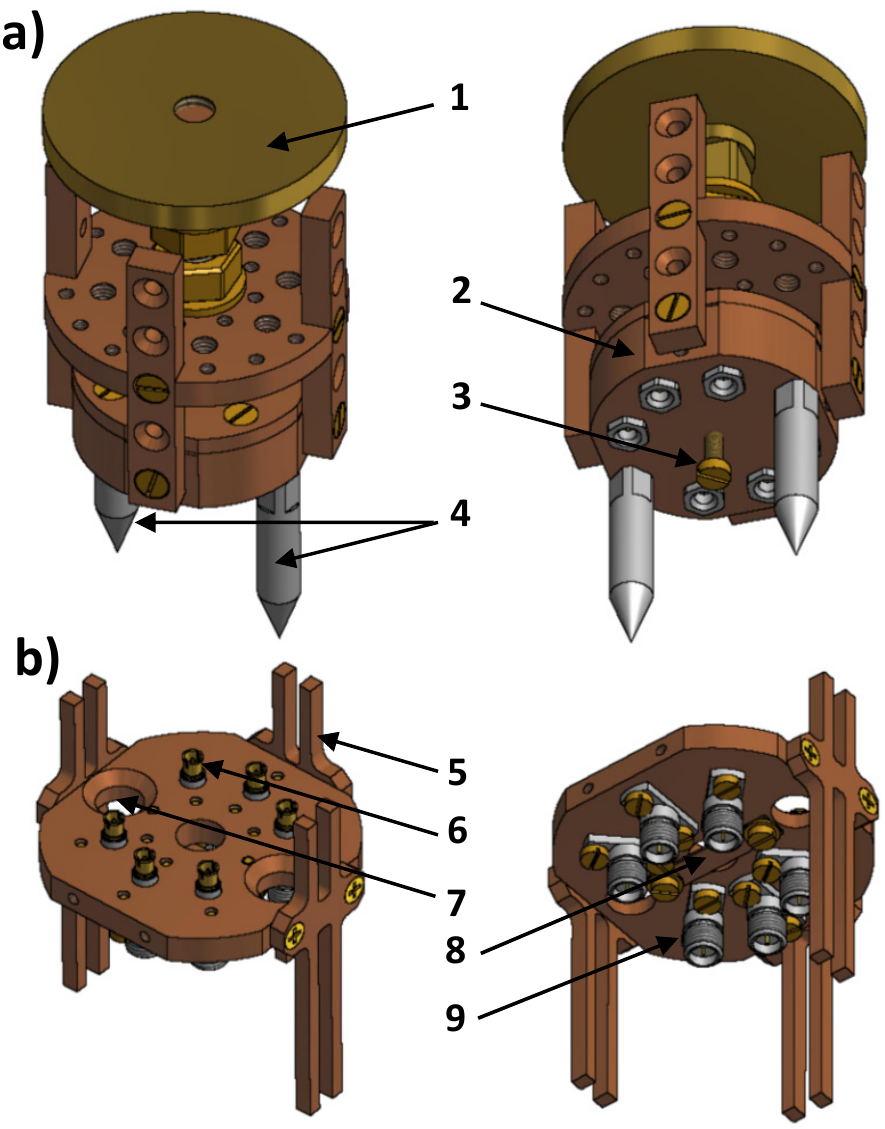}
 \caption{
   \label{fig2}
Drawings of the mechanical assemblies, seen from two point of views. The top assembly (a) is attached to the end of the insertable probe, while the bottom assembly (b) is permanently fixed to the mixing chamber of the refrigerator. 1: cold end of probe. 2: sample holder. 3: stopper. 4: teflon guiding rods. 5: rails, height adjustable with respect to the mixing chamber. 6: SMP-SMP adapter ("bullet"). 7: guiding hole. 8: berylium copper strip. 9: SMP-SMA adapter.}%
 \end{figure}

\begin{figure}[htbp]
\includegraphics[scale=0.5]{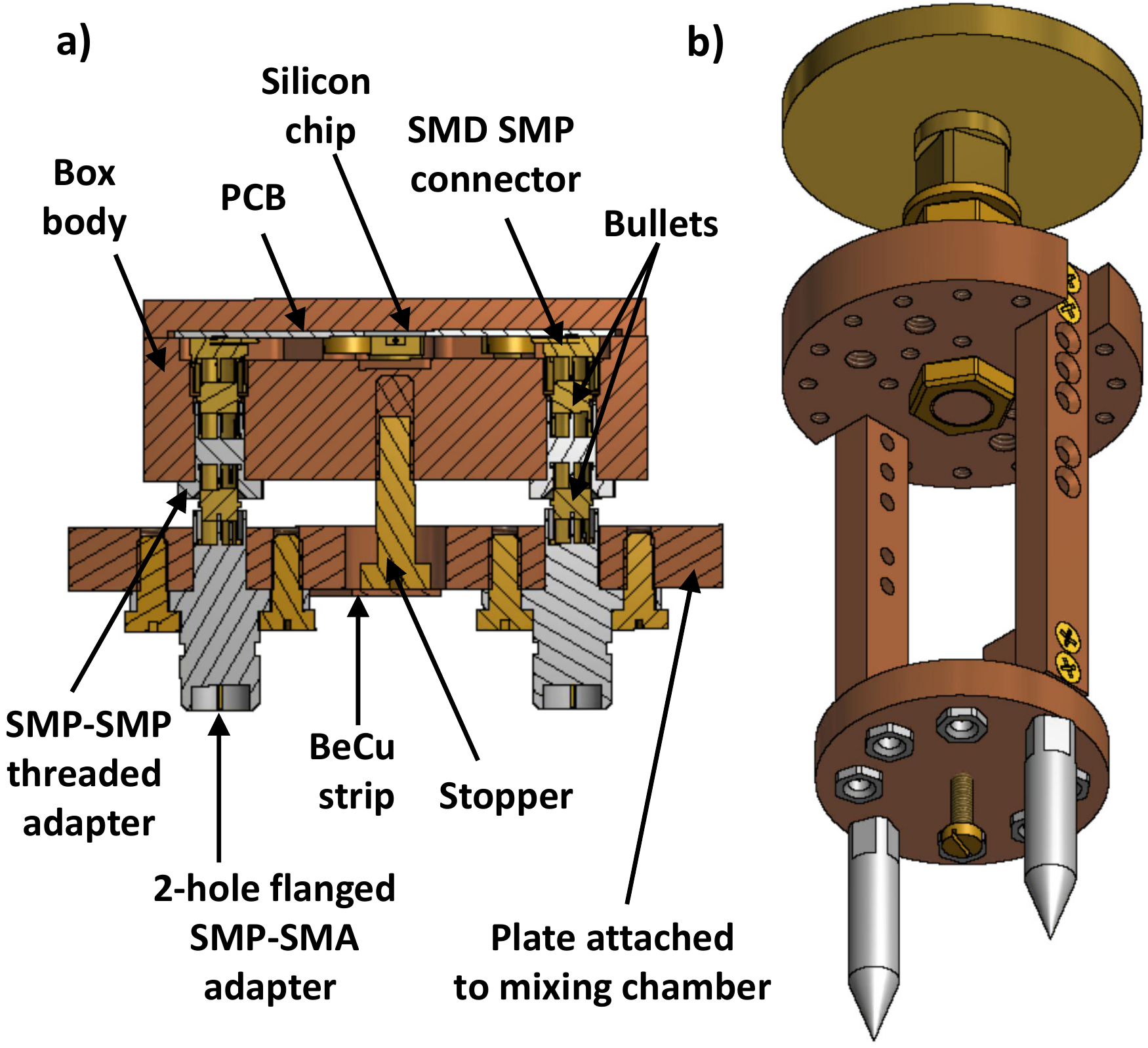}
 \caption{
  \label{fig3}
 Interfacing between insertable devices and the microwave setup. a) Close-up section of the interconnected part of the setup presented on Fig.\ref{fig2}, in the situation where top and bottom assemblies mate. b) Alternative to assembly used to connect sample holders not specifically designed to be mated directly onto the bottom assembly.}%
 \end{figure}

Figure~\ref{fig2} shows a drawing of the insertion system, consisting of two mechanical assemblies. The top assembly is attached to the cold end of the insertable probe (Fig. \ref{fig1}.b) and thus is mobile. It holds the sample holder to be cooled down. The bottom assembly is attached to the mixing chamber of the refrigerator. The top and bottom assemblies are electrically interconnected by an arrangement of SMP adapters\cite{MW-components} (cf Fig.\ref{fig3}.a) including spring connectors (so-called bullets). The latter allow for slight misalignment without impairing microwave properties. All the metallic parts are machined from Oxygen-Free High-Conductivity (OFHC) copper for optimal thermalization. The temperature of the inserted device is measured using a calibrated 100 Ohm SPEER carbon resistor thermometer thermally anchored to the cold end of the probe. The sample holder and its fixture to the cold end are machined out of OFHC copper, ensuring good thermalization.

The transmission lines used in the experiments run from the top plate of the refrigerator (where they are fed into the IVC using vacuum feedthrougs) down to the mixing chamber plate. They are mechanically attached to all the refrigerator plates for proper thermal anchoring. The lines are terminated by SMA microwave connectors plugged to the bottom mechanical assembly.

The top mechanical assembly is attached to the lowest stage of the cold-insertable plate, which is thermally clamped to the mixing chamber after the probe is fully inserted. This assembly carries the sample holder. The substrate (typically a silicon chip) which contains the device to be measured is connected by wire bondings to a printed circuit board (PCB) (Fig. \ref{fig3}.a). Both the chip and the PCB are enclosed in the sample holder. The free volume inside the sample holder is kept as small as possible to move any parasitic resonance above our measurement bandwidth (typically 20 GHz). Launching SMP connectors are soldered on the PCB. Microwave connections to the outside of the sample holder are realized with threaded SMP-SMP adapters which prevent RF leakage in and out the device's space.

The threaded SMP adapters directly mate with bullet adapters on the bottom assembly. The bullets are designed to accommodate relatively large axial and longitudinal misalignements without a significant degradation of the microwave transmission up to 40 GHz. This is important for this system where significant misalignement may occur during insertion and cooldown.

The alignment of the two assemblies prior to connector mating is done using two teflon rods (diameter 7 mm) in the top assembly sliding through guiding holes (diameter 8 mm) in the bottom assembly. The holes are tapered and the teflon rods are terminated in a conical profile to enable easy reach. Coarse angular alignment is done simply using visual marks on the room temperature parts of the probe. A stopper system is designed with a screw protruding from the top assembly, whose head lies on a beryllium copper strip attached to the bottom assembly. This arrangement prevents all the weight of the probe from being entirely supported by the RF connectors when the probe is in the fully inserted position.

Magnetic shielding of the experiment, which is very critical \emph{eg} for work with superconducting flux qubits or squid devices, is done in the following way. A magnetic shielding system, formed of three concentric cylindrical high-magnetic permeability layers, is permanently attached to the mixing chamber plate of the dilution refrigerator. This shield surrounds the bottom mechanical assembly and enables reaching shielding factors between typically 100 and 1000, depending on the insertion depth. This configuration allows for good magnetic shielding without requiring the shielding element to be part of the sample holder, which simplifies the setup.

We note that the connection between the fixed RF lines and the sample can be made in a more indirect but more versatile way than presented above, where the sample holder had to be specifically designed to mate the bottom assembly. Fig.\ref{fig3}.b shows the top assembly of this alternative arrangement (the bottom assembly is the same as above). The sample holder of Fig.\ref{fig2} and Fig.\ref{fig3}.a is replaced here by a simple plate hosting threaded SMP-SMP adapters mating the bottom assembly. In contrast with the previous arrangement the upper side of these adapters is now free to be connected via coaxial cables to any sample holder that fits in the experimental space. This configuration adds more flexibility to the setup, and allows for wiring sample holders whose design prevents direct connector mating. In the following we will focus our discussion on the direct configuration (Figs. \ref{fig2} and \ref{fig3}.a), but the results hold for the alternative configuration as well.

\section{Operation}
    \label{section-operation}

We describe in this section the experimental procedure to cooldown and connect a device with the insertable probe, indicating the duration of each step.

We start in a state where the IVC contains exchange gas and all the plates of the refrigerator have a temperature of approximately 4 K. This temperature is maintained by running the pulse tube cooler. The dilution circuit is under vacuum and the $^3$He-$^4$He mixture is stored at room temperature. The IVC port used for probe insertion (in this case the central 50 mm line of sight port) is isolated by a manual gate valve.

Once the device is wire-bonded to the PCB and enclosed in the sample holder, we attach the top assembly to the end plate of the probe using a threaded rod. The probe is attached to the IVC port and its loadlock is pumped for typically 30 min to reach a few $10^{-2}$ mBar. Then the gate valve to the IVC is opened, the sliding seal is loosened, and the probe is inserted in the IVC. Visual markers on the load lock are used to roughly align the teflon guides of the top assembly with the guiding holes of the bottom assembly, before inserting the probe all the way down and mating the RF connectors. Finally the sliding seal is tightened and the probe is brought in thermal contact with the refrigerator plates. Note that since room temperature parts are brought in contact with the plates at $\approx$ 4 K, the insertion procedure has to be done slowly and carefully, which takes approximately 15 min.

Due to the considerable heat transferred to the refrigerator the temperatures of the cold plates increase up to approximately 50 K. This heat is extracted by the pulse tube cooler. After 3 hours all the IVC volume is thermalized at $\approx$ 3.8 K. Once the exchange gas in the IVC has been adsorbed by charcoals ($\approx 10$ min) the mixture can be condensed. After an additional 2h30 the coldest stage of the probe reaches 80 mK. 2h30 are further needed to reach a temperature of 40 mK. In stationary regime, the probe temperature settles typically to 10 mK above the dilution refrigerator base temperature\cite{note-thermalization}.

Removing the probe proceeds as follows. First the $^3$He-$^4$He mixture is recovered. This step is optional, however it has the following advantages: the risk of sudden pressure increase in the dilution unit is removed and it leaves the system in a state compatible with the next use of the insertable probe. After the thermal clamps are released and the sliding seal slightly loosened, the probe is retracted in the loadlock, the IVC is isolated, and exchange gas is introduced to speed up thermalization to room temperature. The overall time to remove the probe and bring it to room temperature is 1h to 1h30.

In summary it takes less than 6h30 between sample mounting and performing measurements at 80 mK, which is a temperature low enough to characterize superconducting devices involving aluminium Josephson junctions in a regime where quasiparticle poisoning is negligible~\cite{tinkham}. The critical time is the thermalization to 4 K after insertion from room temperature, since all the cooling power is provided by a pulse tube rather than Helium vapors as in a regular dip-stick configuration.

At the time we write this manuscript this operating mode has been repeated over 40 times in 8 months and the design has proven extremely robust. Actually not a single mechanical part or RF connector has needed replacement yet. As discussed in detail in the next section, the RF properties have also proven reliable and stable over time.

We note that it is possible to insert/remove the probe without extracting the mixture. Maintaining the circulation has the advantage that the temperature increase during probe insertion/removal is significantly lower, which is advantageous if other experiments are being done on devices attached to the refrigerator mixing chamber plate. The disadvantage of this alternative method is the fact that probe insertion requires a more controlled thermalization procedure, by successively clamping to all the plates during insertion, which requires significantly more care. In addition, there is a more significant risk of uncontrolled pressure increase in the dilution circuit.

\section{Examples of measurements}

In this section we present examples of microwave measurements performed with the insertable system. We focus here on experiments related to the field of superconducting quantum devices, however we emphasize that this approach can be applied to any experiment requiring microwave frequencies and dilution temperatures. All the RF components used to build the mobile connections are rated up to at least 40 GHz. However the fixed parts of the measurement lines (SMA connectors) are specified to 18 GHz, and our microwave setup contains circulators whose maximum working frequency is 12 GHz. So in this work the tested bandwidth is DC to 12 GHz, but we expect the performances of the design to be similar up to at least 18 GHz.

We first present the characterization of the assembly by measuring the transmission using a wideband through connection. Then we present transmission measurements of a coplanar waveguide (CPW) resonator. Finally we show measurements performed on a superconducting qubit coupled to a microwave resonator.

\subsection{Measurement of transmission using a through transmission line}
\label{through-meas}

A first characterization of the setup consists in measuring the microwave transmission of the RF interconnections, to ensure that the stacking up of RF connectors described in Section \ref{section-setup} does not introduce unacceptable insertion loss or spurious resonances. For that purpose a PCB which contains a coplanar wave guide (CPW) is mounted in the sample holder.

\begin{figure}[htbp]
\includegraphics[scale=0.5]{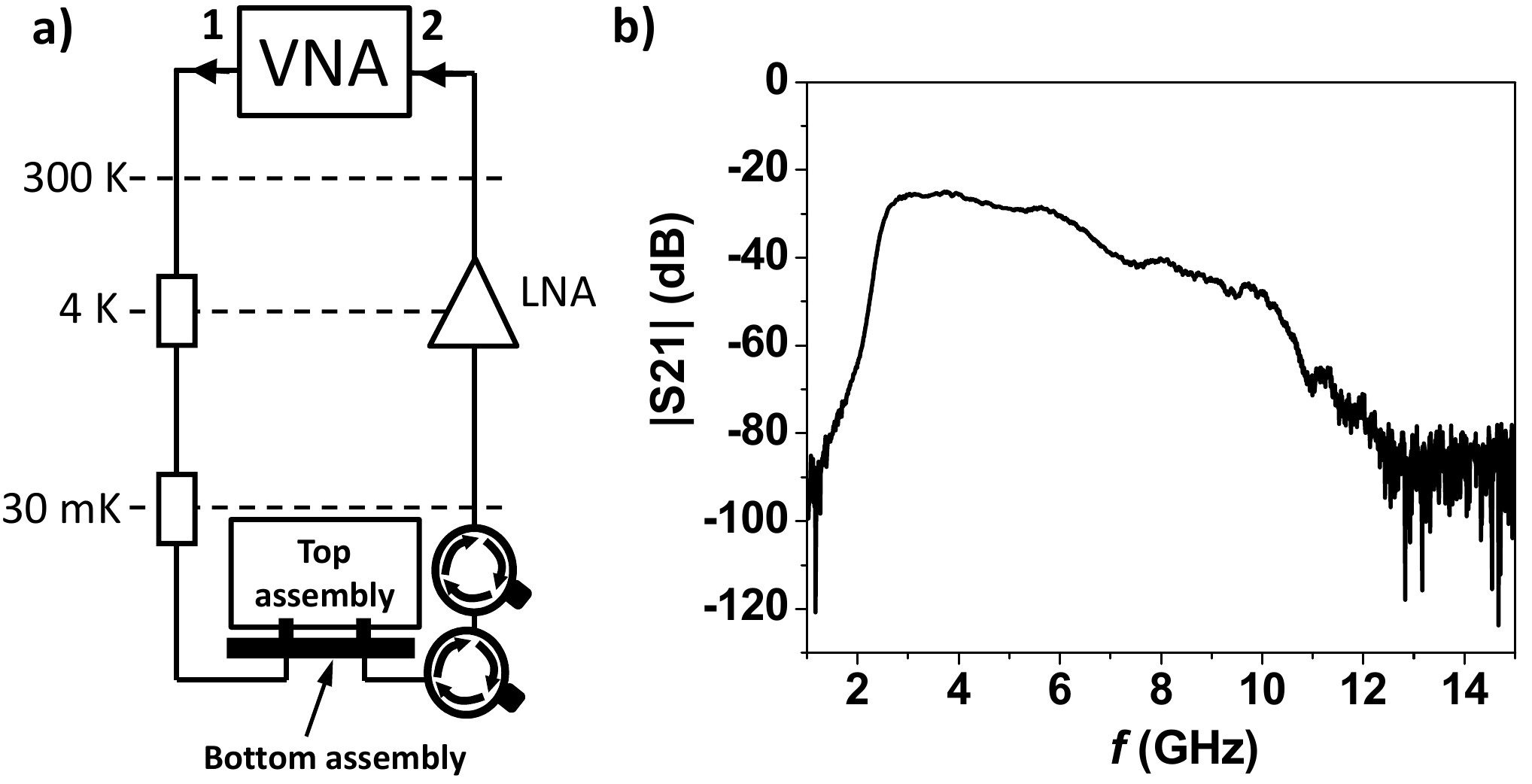}
 \caption{
  \label{fig4}
 Characterization of the assembly. a) Setup used to measure the microwave transmission of a device. The rectangles are attenuators, the circles with arrows are isolators. LNA = Low Noise Amplifier. b) Modulus of the transmission through a coplanar waveguide PCB at $T=$ 50 mK.}
 \end{figure}

The measurement setup is sketched on Fig.\ref{fig4}.a. A vector network analyser (VNA) sends a microwave tone of frequency $f$ from its Port 1. The signal travels through coaxial lines and attenuators down to the bottom assembly where it is connected to the sample holder. The output signal passes through two isolators, is amplified by a cryogenic Low Noise Amplifier (LNA), and reaches Port 2 of the VNA. The VNA measures the complex transmission $S_{21}$ from Port 1 to Port 2 as a function of $f$.

In a preliminary characterization at room temperature, with access to the mixing chamber, we measured the transmission as sketched on Fig.\ref{fig4}.a and compared it to a reference measurement where the whole assembly was replaced with a coaxial cable (data not shown). In the former case $|S_{21}(f)|$ lies 0.1 to 1 dB below the reference and does not exhibit sudden variations or modulations. This shows that the interconnecting scheme of the assembly does not act as a significant source of loss or reflection at room temperature.

Fig.\ref{fig4}.b shows the amplitude of $S_{21}$ measured at $T=$ 50 mK after the insertion procedure described in Section \ref{section-operation}. $S_{21}$ is a smooth function of $f$, which is remarkable given the multiple interconnections involved in the setup. By calculating the attenuation and gain of the measurement line at a few frequencies spanning our measurement bandwidth (2-10 GHz), we estimate the losses added by the insertable assembly to lie within the uncertainty range ($\pm$ 2 dB) of the overall expected transmission.

The connecting of the microwave connectors is very reproducible and was tested by applying the following protocol: connect the coplanar wave guide at low temperature, measure transmission, release, warm up, cool down, reconnect, remeasure. Within the uncertainty of the VNA (0.2 dB) we observe no difference on the transmission between two successive measurements of the same device.

\subsection{Measurement of a coplanar waveguide resonator}

\begin{figure}[htbp]
\includegraphics[scale=0.7]{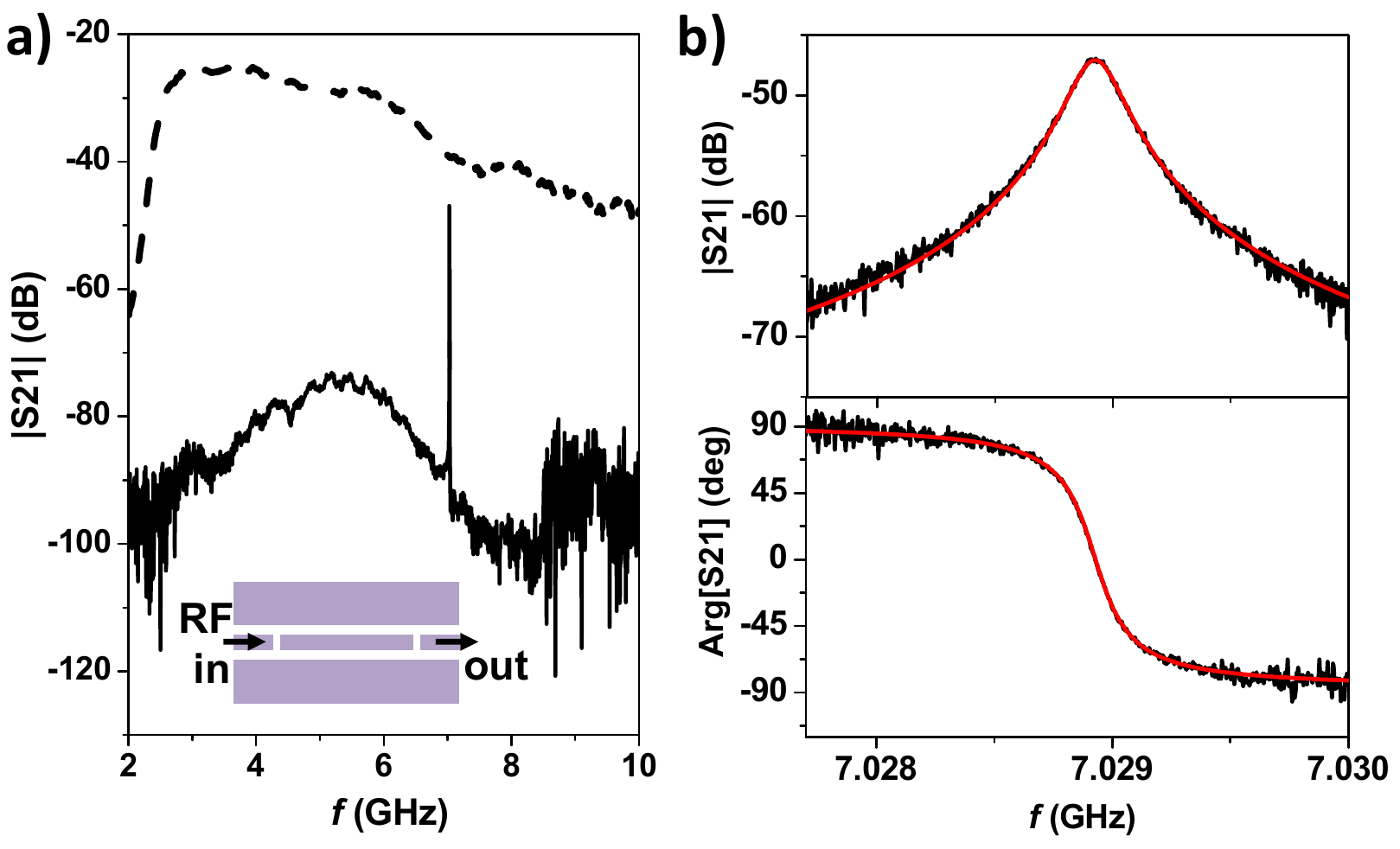}
 \caption{
  \label{fig5}
 Transmission measurement of a CPW resonator. a) Full line: amplitude of $S_{21}$ over the full measurement bandwidth. Dashed line: reference transmission measured with a through PCB. Inset: sketch of the device. b) Close-up view of the complex transmission on the resonance probed at the single photon level. The red thick lines show Eq.\ref{S21-trans} for the best fitted values of circuit parameters.}
 \end{figure}

 We now turn to the characterization of a superconducting microwave resonator, a model system in microwave engineering \cite{pozar} as well as a building block for various areas of physics, including photon detection devices for astronomy \cite{zmuidzinas2003}, circuit quantum electrodynamics (circuit QED) \cite{blais2004,wallraff2004}, or quantum limited amplifiers \cite{clerk2010,lehnert2008,devoret-JPC2010}. The device presented here is a distributed element resonator made of a CPW whose central line is interrupted by two gaps forming capacitors and defining a cavity \cite{pozar}, as sketched in the insert of Fig.\ref{fig5}.a. The device is made from aluminium (thickness 200 nm) on a silicon substrate, using a liftoff process.

The measurement setup is identical to the one used in \ref{through-meas} and sketched on Fig.\ref{fig4}.a. The complex transmission $S_{21}$ for this device is shown in Fig.\ref{fig5}.a over the full measurement bandwidth (full line). We observe a transmission peak around 7 GHz corresponding to the first mode of the resonator. Apart from this peak, $|S_{21}|$ lies well below the transmission measured previously with a through PCB and reproduced in Fig.\ref{fig5}.a as the dashed line. In particular $|S_{21}|$ does not exhibit parasitic features like box resonances or spurious capacitive coupling between ports. The absence of parasitic transmission combined to the insertion loss below 2 dB stated previously can be seen as a proof of the robustness of the design from the microwave point of view.

We show on Fig.\ref{fig5}.b a close-up view of the first mode probed at the level of one photon populating the cavity on average when the drive is resonant. To separate the external and internal quality factors (respectively $Q_{\rm e}$ and $Q_{\rm i}$, yielding a total quality factor $Q_{\rm t} = (1/Q_{\rm e} + 1/Q_{\rm i})^{-1}$) we fit the complex transmission $S_{21}(f)$ with the transfer function:

\begin{equation}
\tau(f) = \frac{A}{Q_{\rm e}} \frac{1}{ \frac{1}{Q_{\rm t}} + 2 j \frac{f-f_0}{f_0} } e^{j \varphi_0}
\label{S21-trans}
\end{equation}

 where $A=1/|S_{21}^{\rm ref}|$ is a normalization factor obtained directly from the through measurement (dotted line), $f_0$ is the loaded resonance frequency, and $\varphi_0$ is a global phase factor. We obtain $f_0$ = 7.029 GHz, $Q_{\rm e}$ = 53,000 and $Q_{\rm i}$ = 77,000. $f_0$ and $Q_{\rm e}$ are in good agreement with the designed values, and the internal quality factor $Q_{\rm i}$ reaches a state of the art value for non epitaxial aluminium on silicon resonators at the single photon level \cite{sage2011}.

In this subsection we only presented measurements of a CPW type resonator measured in a transmission configuration. However, the versatility of the setup allows for other types of resonators (eg lumped elements circuits, multiplexed notch resonators) measured in either reflexion or transmission configuration.

\subsection{Measurement of a circuit-QED device using a superconducting flux qubit}

\begin{figure}[htbp]
\includegraphics[scale=0.8]{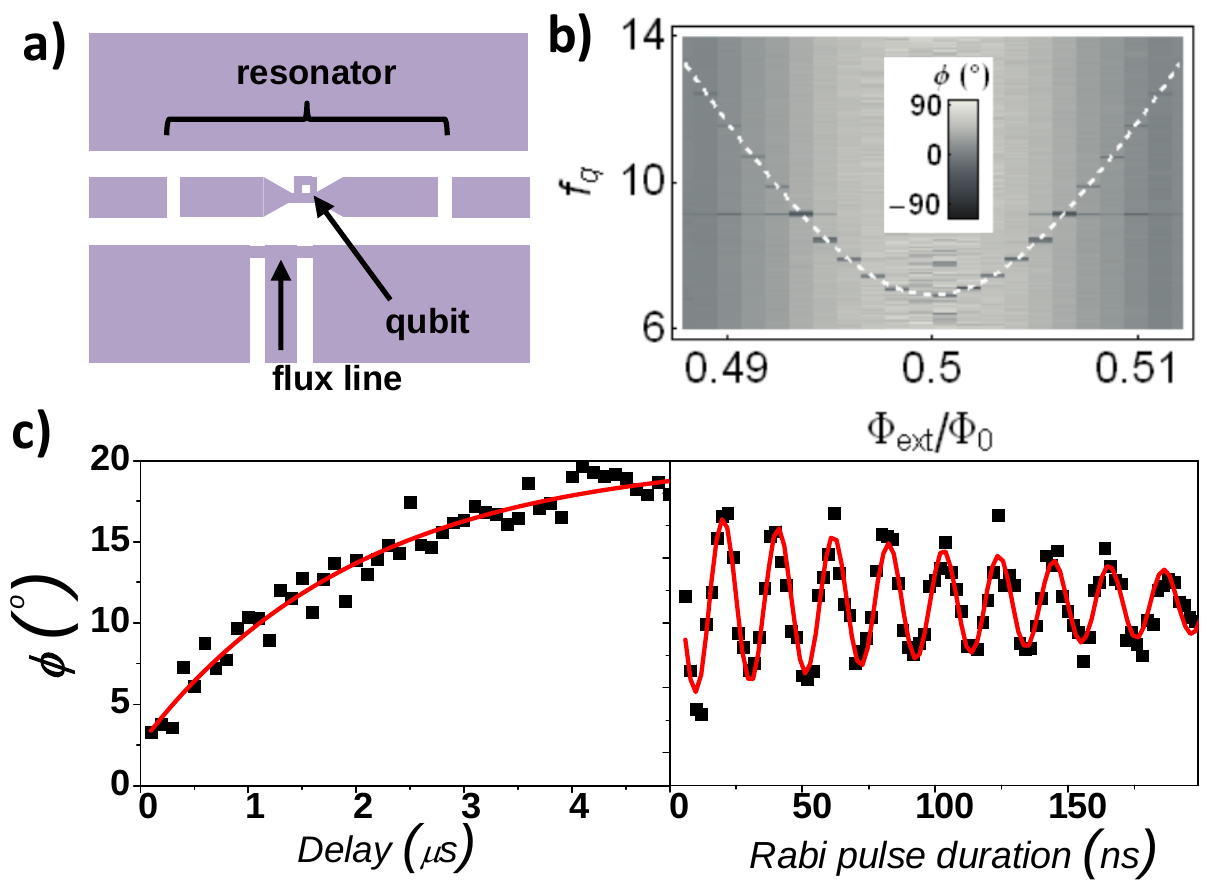}
 \caption{
  \label{fig6}
 Circuit-QED experiment a) Sketch of the device. b) Spectroscopy of the qubit. The dashed line is a fit of the qubit transition frequency. c) Relaxation (left) and Rabi oscillations (right) measurements. Dots are data, red lines are fits.}
 \end{figure}

We conclude this section with results obtained on a circuit-QED experiment \cite{blais2004,wallraff2004} with a flux qubit. The device is sketched on Fig.\ref{fig6}.a and consists of a flux qubit \cite{mooij99} inductively coupled to a coplanar waveguide resonator \cite{abdumalikov08}. A superconducting coil attached to the sample holder enables biasing of the qubit with a DC flux $\Phi_{\rm ext}$, and a local wideband flux line placed close to the qubit is used for fast bias and drive with RF signals.

We only present here the main results and focus on typical figures of merit evaluating the quality of the electromagnetic environment that couples to the device in our setup. For the details we refer the reader to the literature covering the topic \cite{haroche-book,blais2004,wallraff2004}. Circuit-QED focuses on the coherent interaction between light and matter at the single excitation level. Coherence is extremeley sensitive to electromagnetic noise coupling to the device, which can significantly enhance the relaxation and pure dephasing rates of the qubit. Probing the coherence of a qubit is thus a way to measure the electromagnetic isolation of the whole device.

First, qubit spectroscopy is performed as a function of the flux bias applied to the qubit ring. The qubit state is readout using a dispersive measurement scheme \cite{blais2004}: the resonator is driven at its resonance frequency and the qubit-state-dependent phase $\phi$ of the transmitted signal is measured by homodyne detection (Fig. \ref{fig6}.b). The spectroscopic data can be used to extract the parameters describing the flux qubit-resonator coupled system. The dashed line on Fig. \ref{fig6}.b is plotted for a tunneling energy $\Delta= \hbar \times$ 6.90 GHz, a persistent current $I_{\rm p}$ = 150 nA and a qubit-resonator coupling $g = \hbar \times$ 95 MHz.

Next the coherence of the flux qubit is probed. In superconducting qubits, decoherence has two components of comparable importance: relaxation and pure dephasing. Fig. \ref{fig6}.c shows two kinds of time domain experiments allowing to estimate the coherence of a qubit: energy relaxation (left panel), and Rabi oscillations (right panel). Both experiments are performed at $\Phi/\Phi_0=0.507$.
The relaxation is exponential with a time constant $T_1=(2.0 \pm 0.2)$ $\mu$s, whereas the characteristic damping time of the Rabi oscillations is $T_{\rm R} = (180 \pm 25)$ ns. These values of $T_1$ and $T_{\rm R}$ are consistent with state of the art experiments involving flux qubits away from the symmetry point \cite{working-point}.
We thus conclude that our insertable system can be used for sensitive experiments where preserving coherence is challenging.

\section{Conclusion}

We designed and tested a mechanical system for connection of devices to high frequency
transmission lines into a running dilution refrigerator. This system is used in combination
with a cold insertable probe to perform fast turnaround experiments using low noise large
bandwidth electrical measurements at temperatures below 100 mK. The total time to
cool a device down to 80 mK and warm it up to room temperature using this system is 8 hours, which is a major improvement
over the regular mode of operation used so far in similar experiments. This system is
robust: it was used over 40 times over a time period of 8 months. We showed two examples
of measurements enabled by this method. The first example is a transmission measurement of
a superconducting cavity. The second experiment is a study of a persistent current qubit,
in which the qubit has coherence time comparable with state of the art at this time in the
field. This method will be highly relevant to various experiments in the field of low-temperature physics, in particular for quantum coherence studies.


%
%

\begin{acknowledgments}
We acknowledge Harmen Vander Heide, Andrew Dube and Michael Lang from Science Technical Services at University of Waterloo for help with design and realization of the mechanical assembly. We also acknowledge Mustafa Bal and Chunqing Deng for acquiring the data on qubit measurements using this setup. This work was supported by the Natural Sciences and Engineering Research Council of Canada, the Waterloo Institute for Nanotechnology, and the Alfred Sloan Fundation. The infrastructure used for this work would not be possible without the significant contributions of the Canada Foundation for Innovation, the Ontario Ministry of Research and Innovation and Industry Canada. Their support is gratefully acknowledged.
\end{acknowledgments}


\begin{thebibliography}{29}
\bibitem [{MW-()}]{MW-components}%
  \BibitemOpen
  \href@noop {} {}\bibinfo {note} {All the SMP connectors used in this work are
  available either from Rosenberger (part numbers 19S102-40ML5 and 19K101-K00) or Fairview Microwave (part numbers SM8843, SM2614 and SM8794).}\BibitemShut {Stop}%
\bibitem [{wor()}]{working-point}%
  \BibitemOpen
  \href@noop {} {}\bibinfo {note} {The coherence is expected to be optimal at
  $\Phi/\Phi_0=0.5$ (symmetry point). However at that working point the
  detuning between the flux qubit and the resonator is not large enough
  compared to $g$ so their dynamics are hybridized. To be sensitive to the
  coherence of the qubit only we chose to work away from the symmetry
  point.}\BibitemShut {Stop}%
\end{thebibliography}

%

\end{document}